\documentclass[conference]{IEEEtran}
\IEEEoverridecommandlockouts

\usepackage[margin=1in]{geometry}
\usepackage[utf8]{inputenc}
\usepackage{graphicx}
\usepackage[draft]{hyperref}
\usepackage{amsmath}
\usepackage{amssymb}
\usepackage[acronym]{glossaries}
\usepackage{multirow}
\usepackage{stackengine,scalerel}
\usepackage{accents}
\usepackage{algorithm}
\usepackage{algorithmic}
\usepackage{gensymb}
\usepackage{cite}
\usepackage{amsmath,amssymb,amsfonts}
\usepackage{graphicx}
\usepackage{textcomp}
\usepackage{xcolor}
\usepackage{optidef}
\usepackage{multirow}
\usepackage{stfloats}
\usepackage{lipsum}
\usepackage{subcaption}
\usepackage{optidef}
\usepackage{titlesec}
\usepackage{subcaption}
\usepackage{enumitem}
\glsdisablehyper   

\newacronym[]{AoD}{AoD}{angle of departure}
\newacronym[]{AoA}{AoA}{angle of arrival}
\newacronym{ADC}{ADC}{Analog-to-Digital Converter}
\newacronym{AO}{AO}{alternating optimization}
\newacronym{BB}{BB}{baseband}
\newacronym[]{BF}{BF}{beamforming}
\newacronym{BS}{BS}{base station}
\newacronym[]{CRB}{CRB}{Cramer-Rao bound}
\newacronym{CRSM}{CRSM}{cluster ray sensing model}
\newacronym{DL}{DL}{downlink}
\newacronym{DoA}{DoA}{direction of arrival}
\newacronym{DoD}{DoD}{direction of departure}
\newacronym{DFT}{DFT}{discrete Fourier transform}
\newacronym{IDFT}{IDFT}{inverse discrete Fourier transform}
\newacronym{DFRC}{DFRC}{dual-function radar-communication}
\newacronym{FD-ISAC}{FD-ISAC}{full duplex integrated sensing and communication}
\newacronym{FOV}{FOV}{field of view}
\newacronym{5G}{5G}{fifth generation}
\newacronym{6G}{6G}{sixth generation}
\newacronym{FD}{FD}{full duplex}
\newacronym{FDD}{FDD}{frequency division duplex}
\newacronym{FMCW}{FMCW}{frequency-modulated continuous-wave}
\newacronym{HBF}{HBF}{Hybrid beamforming} 
\newacronym{HD}{HD}{half duplex}
\newacronym[]{ISAC}{ISAC}{integrated sensing and communication}
\newacronym{IID}{IID}{independent and identically distributed}
\newacronym{LOS}{LOS}{line of sight}
\newacronym{MUSIC}{MUSIC}{MUltiple SIgnal Classification}
\newacronym[]{MPC}{MPC}{multi path component}
\newacronym[]{MIMO}{MIMO}{multiple input multiple output}
\newacronym[]{mmWave}{mmWave}{millimeter-Wave}
\newacronym{MSS}{MSS}{maximum signal strength}
\newacronym{MSE}{MSE}{mean square error}
\newacronym{NLOS}{NLOS}{non line of sight}
\newacronym{NSP}{NSP}{null space projection}
\newacronym{NR}{NR}{new radio}
\newacronym{OFDM}{OFDM}{orthogonal frequency division multiplexing}
\newacronym{PAPR}{PAPR}{peak-to-average-power-ratio}
\newacronym{PSD}{PSD}{positive semi-definite}
\newacronym{QoS}{QoS}{quality-of-service}
\newacronym{RSU}{RSU}{road side unit}
\newacronym[]{RX}{RX}{receiver}
\newacronym[]{SINR}{SINR}{signal-to-interference plus noise ratio}
\newacronym{SI}{SI}{self-interference}
\newacronym{SNR}{$\textrm{SNR}$}{signal to noise ratio}
\newacronym{SDR}{SDR}{semi-definite relaxation}
\newacronym{SRSM}{SRSM}{single ray sensing model}
\newacronym[]{TX}{TX}{transmitter}
\newacronym{ULA}{ULA}{uniform linear array}
\newacronym{USA}{USA}{uniform symmetric array}
\newacronym{UPA}{UPA}{uniform planar array}
\newacronym[]{UE}{UE}{user equipment}
\newacronym{UL}{UL}{uplink}
\newacronym{V2I}{V2I}{vehicle-to-infrastructure}
\newacronym{V2X}{V2X}{vehicle-to-everything}
\newacronym{WP2}{WP2}{Work Package 2}
\newacronym{ETSI}{ETSI}{European Telecommunications Standards Institute}
\newacronym{3GPP}{3GPP}{$3^{\rm rd}$ Generation Partnership Project}
\newacronym{RCS}{RCS}{radar cross section}
\newacronym{RV}{RV}{random variable}
\newacronym{TDD}{TDD}{time division duplex}
\newacronym{SVD}{SVD}{singular value decomposition}
\newacronym{SOCP}{SOCP}{second order cone programming}
\newacronym{SOC}{SOC}{second order cone}
\newacronym{CR}{CR}{cluster ray}
\newacronym{TMS}{TMS}{truncated MUSIC spread}
\newacronym{CMS}{CMS}{conventional MUSIC spread}
\newacronym{RMSE}{$\textrm{RMSE}$}{root mean squared error}
\newacronym{DISPARE}{DISPARE}{DOA estimation of distributed signals}
\newacronym{ET}{ET}{extended target}
\newacronym{TFS}{TFS}{truncated Fourier series}
\newacronym{SOTA}{SOTA}{state of the art}
\newacronym{MST}{MST}{manifold separation technique}
\newacronym{FFT}{FFT}{fast Fourier transform}
\newacronym{IFFT}{IFFT}{inverse fast Fourier transform}

\renewcommand{\a}{\mathbf{a}}

\newcommand{\e}{\mathbf{e}}

\newcommand{\h}{\mathbf{h}}

\renewcommand{\r}{\mathbf{r}}
\newcommand{\s}{\mathbf{s}}

\renewcommand{\v}{\mathbf{v}}
\newcommand{\w}{\mathbf{w}}
\newcommand{\x}{\mathbf{x}}
\newcommand{\y}{\mathbf{y}}
\newcommand{\z}{\mathbf{z}}

\newcommand{\0}{\mathbf{0}}

\newcommand{\A}{\mathbf{A}}
\newcommand{\B}{\mathbf{B}}
\newcommand{\C}{\mathbf{C}}
\newcommand{\D}{\mathbf{D}}
\newcommand{\E}{\mathbf{E}}

\newcommand{\G}{\mathbf{G}}
\renewcommand{\H}{\mathbf{H}}
\newcommand{\I}{\mathbf{I}}
\newcommand{\J}{\mathbf{J}}

\newcommand{\Q}{\mathbf{Q}}
\newcommand{\R}{\mathbf{R}}

\newcommand{\V}{\mathbf{V}}

\newcommand{\Y}{\mathbf{Y}}
\newcommand{\Z}{\mathbf{Z}}

\newcommand{\etav}{\boldsymbol{\eta}}
\newcommand{\thetav}{\boldsymbol{\theta}}

\newcommand{\sigmav}{\boldsymbol{\sigma}}

\newcommand{\Gammab}{\mathbf{\Gamma}}

\newcommand{\Lambdab}{\mathbf{\Lambda}}

\newcommand{\Upsilonb}{\boldsymbol{\Upsilon}}

\newcommand{\Psib}{\mathbf{\Psi}}

\newcommand{\setK}{\mathcal{K}}

\newcommand{\diag}{\mathrm{diag}}

\newcommand{\herm}{\mathrm{H}}

\newcommand{\tran}{\mathrm{T}}

\renewcommand{\rm}[1]{\mathrm{#1}}
\newcommand{\norm}[1]{\left\lvert\left\lvert #1 \right\rvert\right\rvert}
\newcommand{\abs}[1]{\left\lvert #1 \right\rvert}

\newcommand{\qed}{\nobreak \ifvmode \relax \else
      \ifdim\lastskip<1.5em \hskip-\lastskip
      \hskip1.5em plus0em minus0.5em \fi \nobreak
      \vrule height0.75em width0.5em depth0.25em\fi}

\allowdisplaybreaks
\titlespacing*{\subsection}
  {0pt}{1\baselineskip}{1\baselineskip}

\titlespacing*{\subsection}
  {0pt}{1\baselineskip}{1\baselineskip}

\newcommand{\RN}[1]{%
  \textup{\uppercase\expandafter{\romannumeral#1}}%
}
\def\BibTeX{{\rm B\kern-.05em{\sc i\kern-.025em b}\kern-.08em
    T\kern-.1667em\lower.7ex\hbox{E}\kern-.125emX}}

\begin{document}


\title{{Full Duplex ISAC with Cluster Ray Targets: Parameter Estimation and Beamforming}
\thanks{This project is funded by Continental Automotive Technologies GmbH under grant DG-088181.}
}

\author{\IEEEauthorblockN{ Muhammad Talha}
\IEEEauthorblockA{\textit{Electrical and Computer Engineering} \\
\textit{University of Illinois Chicago, USA}\\
mtalha7@uic.edu}
\and
\IEEEauthorblockN{ Besma Smida}
\IEEEauthorblockA{\textit{Electrical and Computer Engineering} \\
\textit{University of Illinois Chicago, USA}\\
smida@uic.edu}
\and
\IEEEauthorblockN{David Gonz\'{a}lez G.}
\IEEEauthorblockA{\textit{Continental Technologies GmbH} \\
Germany \\
david.gonzalez.g@ieee.org}
}
\setlength{\abovedisplayskip}{5pt}
\setlength{\belowdisplayskip}{5pt}
\maketitle
\vspace{-1cm}
\begin{abstract}
This work studies a full-duplex integrated sensing and communication (ISAC) resolution framework for spatially distributed systems. Conventional high-resolution methods, such as MUSIC, fail to localize distributed targets because the signal subspace is full rank, even in the single-distributed-target setting. In an effort to resolve this, we propose a two-stage estimator, which successfully resolve multiple distributed targets and outperforms several baseline schemes without incurring any additional computational complexity. Our first-stage estimator uses the Fast Fourier transform to estimate the coarse spectrum, while in the second stage, we apply the Gauss-Newton method to fine-tune the angular estimates. Apart from this, we also propose an optimization framework for designing an adaptive beamformer capable of synthesizing both wide and directed beams to cover the full extent of the targets while also fulfilling data rate requirements of multiple users. The beamformer also meets the data-rate requirements of multiple users, maintaining quality of service. Simulation results demonstrate a threefold improvement in spread estimation under low signal-to-noise ratio (SNR) conditions and a twofold improvement for low-spread targets.

\end{abstract}
\section{Introduction}
\Gls{ISAC} systems have been one of the most important areas of research for next-generation \gls{6G} wireless systems \cite{FeasibilityStudy3gpp}. In \gls{ISAC} systems, various taxonomies have been investigated in the literature based on system models and waveform designs \cite{ISAC6GSurvey}. For instance, from a system modeling perspective, \textit{device-free} and \textit{device-based} sensing scenarios have been investigated \cite{FundamentalLimitsSurvey}, while in the waveform design aspect, \textit{communication-centric, radar-centric, and joint-design} strategies have been explored and advanced in recent years \cite{majorcomm}.

Despite the recent advancements in \gls{ISAC} literature, most of the works do not focus on the modeling aspect of radar targets \cite{BF_Design_FD_ISAC2022,OptimalTXBFforISAC}. For instance, in most \textit{communication-centric} design studies, both sensing targets and communication users have been modeled as point objects in the far field of \gls{TX} antennas \cite{JTRBFFDISAC,TalhaAsilomar2023,ISACATIQ}. This modeling aspect is acceptable for communication purposes as the \gls{RX} antennas occupy a significantly small area; however, for sensing in \gls{V2I} and \gls{V2X} settings, this assumption becomes invalid.
Realizing this, many research groups have focused their attention on the \gls{ET} model\cite{ISACExtendedTarget,crboptimizationJRCB}, or on using a parametric model to define a contour of the target, such as \gls{TFS} \cite{ExtendedtargetTFS,3DextendedTargetTFS}. Although these models can model complex targets, they do so at a much higher computational cost. For instance, in these strategies, either all scatterers or a large number of parameters are required to fully characterize the target. In addition, these models assume deterministic settings and thus ignore the stochastic nature of the radar channel. 

In an effort to address these limitations, we use a parametric distributed \gls{CR} target model, in which each target is modeled as a cluster of rays with corresponding densities. Based on this, the main contributions of our work are:
\begin{itemize}[leftmargin = 2mm]
    \item Realizing the computational complexities of target modeling in recent works \cite{ExtendedtargetTFS,3DextendedTargetTFS}, we use a computationally efficient distributed \gls{CR} model for a target --  proposed in \gls{3GPP} \cite{3GPP38901}. To the best of our knowledge, this is the first attempt to study a spatially spread or clustered target in the context of \gls{ISAC} systems. 
    \item We propose a two-stage estimation framework for estimating the angular parameters of multiple targets. In the first stage, we use a \gls{MST}- based approach, which enables construction of the target spectrum using an efficient \gls{FFT} algorithm to linearize the direction parameter in the array signal model. In the second stage, we develop a Gauss-Newton-based approach to fine-tune the coarse estimates. 
    \item We propose a dynamic beam pattern synthesis algorithm that meets multiple users' data rate requirements and limits the \gls{SI} signal power at the receiver RF chains, a consideration most prior works ignore. 
   
\end{itemize}

\section{System Model}
    \begin{figure}[t]
        \centering
\includegraphics[width=1\linewidth]{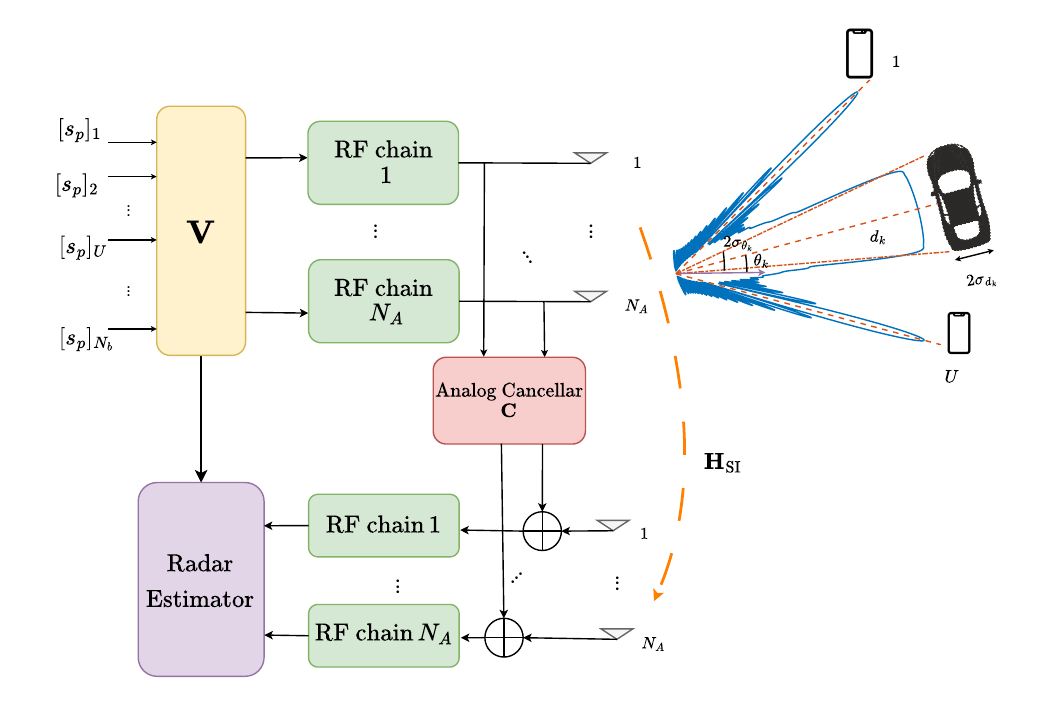}
        \caption{System Model where \gls{FD}-\gls{ISAC} \gls{BS} serves $U$ \gls{DL} communication users along with sensing of $K$ distributed \gls{CR} targets.} 
        \label{fig:system_model}
    \end{figure}
    We consider a \gls{MIMO} \gls{FD} \gls{ISAC} system operating at \gls{mmWave} frequencies using \gls{OFDM} waveforms. We assume that \gls{BS} transmits information over $P$ subcarriers for $U$ users as presented in Fig. \ref{fig:system_model}. \gls{BS} is also equipped with a sensing module capable of estimating the angular parameters from the received reflections from the targets. We assume that \gls{BS} is equipped with $N_A$ transmit and receive antennas and transmits $U$ independent information signals (for $U$ users) as well as $N_A - U$ artificial sensing signals. The information symbol at $p^\mathrm{th}$ subcarrier can then be given as
  $ 
        \s_p = \begin{bmatrix}
            \s_U^\tran & \s_R^\tran
        \end{bmatrix}^\tran \in \mathbb{C}^{N_A \times 1}
    $ 
    where $\s_U \in \mathbb{C}^{U\times 1}$ is intended for $U$ users while $\s_R \in  \mathbb{C}^{N_A - U \times 1}$ is an artificially created radar signal which is uncorrelated to the communication signals i.e., $\mathbb{E}\left(\s_R\s_U^\herm\right) = \0$ and $\mathbb{E}\left(\s_p\s_{p}^\herm\right) = \I $. This information signal then undergoes digital beamforming 
    that maps different streams to different transmit antennas. The overall transmit signal at $p^\mathrm{th}$ subcarrier is defined as
\begin{align}
    \x_p = \V_U\s_U + \V_R\s_R = \V\s_p 
\end{align}
where $\V_U \in \mathbb{C}^{N_A \times U}$ and $\V_R \in \mathbb{C}^{N_A \times N_A - U}$ are beamforming matrices for $U$ users and sensing the \gls{CR} targets. We denote the transmit covariance matrix with $\R_\x = \mathbb{E}\left(\x_p\x_p^\herm\right)$. In addition, we define the set of users as: $\mathcal{L} \triangleq \{1,2,\cdots,U\}$. For target modeling, we denote $N_p$ as the number of rays per cluster, $\alpha_{k,r}$ as the reflection coefficient of the $r^\rm{th}$ ray of the $k^\rm{th}$ cluster. Moreover, we represent the array response vector of direction $\theta$ as $\a_{k,r} = [e^{-j\pi \frac{N_A - 1}{2}\sin(\theta_{k,r})}, \cdots, e^{j\pi \frac{N_A - 1}{2}\sin(\theta_{k,r})} ]^\tran$. The radar channel at $p^\rm{th}$ subcarrier can then be given as
\begin{align}
     \H_\mathrm{rad,p} = \sqrt{\kappa}\sum_{k=1}^K\sum_{r=1}^{N_p}\alpha_{k,r}\omega_{k,r}^{p}\A_{k,r}, \label{eq:RadChannel}
\end{align}
    where $\A_{k,r} \triangleq \a_{k,r}\a^\herm_{k,r}$, $\kappa = \frac{N_A^2}{N_p}$, $\omega_{k,r}^p = e^{-j2\pi p\Delta f \tau_{k,r}}$ denote the phase shift across subcarriers due to the target distance, $\Delta f $ is the subcarrier spacing, $\tau_{k,r} = \frac{2d_{k,r}}{c}$ represent the time delay due to the ray distance $d_{k,r}$ from the \gls{BS} and $c$ denote the speed of light. We assume that the distance follows a certain symmetric probability distribution $\rho_{D_k}(d)$ with a mean distance ($d_{k}$) and range spread ($\sigma_{d_k}$). Moreover, $\theta_{k,r}$ represents the direction of the $r^\rm{th}$ ray of the $k^\rm{th}$ cluster, and it also follows a symmetric spatial distribution $\rho_{\Theta_k}(\theta)$ with mean angle ($\theta_k$) and angular spread ($\sigma_{\theta_k}$). 
    {The signal received at the antenna frontend can be expressed as
    \begin{align}
        \y_p^{\mathrm{RF}} = \H_{\mathrm{rad},p}\x_p + \H_{\mathrm{SI}}\x_p + \z_p
    \end{align}
    where $\H_{\mathrm{SI}}$ is the \gls{SI} channel, which is modeled as Ricean distribution following \cite{SIChannel}. We assume that both analog and digital cancellation architectures are present, and let the analog canceller be denoted by matrix $\C\in \mathbb{C}^{N_A\times N_A}$, and the digital canceller be denoted by $\D\in\mathbb{C}^{N_A\times N_A}$; hence, the signal after analog cancellation is given as
    \begin{align}
        \y_p^{\mathrm{analog}} = \H_{\mathrm{rad},p}\x_p + \H_{\mathrm{SI}}\x_p +\C\x_p + \z_p
    \end{align}

    We design both $\C$ and $\D$ based on the strategy presented in \cite{ISACATIQ}, which assumes that if the \gls{SI} power is below the RF saturation level of \gls{RX} RF chain, then the digital canceller can successfully remove the \gls{SI} signal from the received signal. The post-cancellation signal is given as 
}
    \begin{align}
        \y_p = \H_\rm{rad}^p\x_p + \z_p \label{eq:y_p}
    \end{align}
    where $\z_p \in \mathbb{C}^{N_A}$ is {the residual SI and noise} vector whose each entry is \gls{IID} complex Gaussian with zero mean and variance $\sigma_b^2$, i.e., $\z_p \sim \mathcal{CN} \left(\0,\sigma_b^2 \I_{N_A}\right)$. The \gls{DL} channel of user $u \in \mathcal{L}$ is given as $\h_u^\herm = \sqrt{N_A}\beta_u\a_{\rm{N_A}}^\herm(\phi_u),$ where $\beta_u$ and $\phi_u$ represent the channel coefficient and the LOS direction from the \gls{BS}. The received signal at the $p^\rm{th}$ subcarrier at user $u$ is given as
    \vspace{-2mm}\begin{align}
        r_{p,u} = \h_u^\herm (\v_u s_{p,u} + \sum_{u' \neq u }^{N_A}\v_{u'} s_{p,{u'}}) + z_{p,u},
    \end{align}
    where $\v_u$ is the $u^\rm{th}$ column of $\V$, $s_{p,u}$ is the $u^\rm{th}$ entry of vector $\s_p$, and $z_{p,u}$ represents the noise at the user $u$ following a complex Gaussian distribution with zero mean and variance $\sigma_u^2$. 
    The main objective of the \gls{ISAC} system is to 1) fulfill the data rate requirements of all the $U$ users in the environment, 2) estimate angular parameters, and 3) design a beam pattern to illuminate the full extent of all the targets $K$.

\section{Subspace-based parametric models}
{
Conventional estimation methods for distributed sources \cite{WeightedSubspace,parametricLocalization,CMS} rely on the estimation of the signal and noise subspace of the covariance matrix, i.e., 
\begin{align}
    \mathbf{R}_\y = \sum_{k\in\setK}q_k^2\Psib_k + \sigma_b^2\I \label{eq:Covariance_Matrix},
\end{align}
where $q_k^2 = N_A^2\mathbb{E}[|\alpha_{k,r}|^2]$, and $\Psib_k = \mathbb{E}\left(\A_{k,r}\V\s_p\s_p^\herm\V^\herm\A_{k,r}^\herm\right)$, where the expectation is over the angular distributions and information symbols. For point sources, the rank of each signal matrix $\Psib_k$ is one. However, for distributed targets, these matrices are often full-rank, although most of the energy is encapsulated in the first $q$ eigenvectors \cite{parametricLocalization}. Writing the angular direction as $\theta_{k,r} = \theta_k + \tilde{\theta}_{k,r}$, where $\theta_k$ is the mean angle of target $k$ and $\tilde{\theta}_{k,r}$ is the random error with mean $0$ and spread $\sigma_{\theta_k}$. Given that each target follows common distributions, such as, Uniform, or Gaussian, and each target exhibits small angular spread, i.e., $\sin({\theta}_{k,r}) \approx \sin(\theta_k) + \tilde{\theta}_{k,r}\cos(\theta_k)$, one can find the closed form expression for matrices $\Psib_k$. Using the approximation, the array response vectors can be approximated as $\a_{k,r} \approx \D_k\tilde{\a}_{k,r}$, where $\D_k = \mathrm{diag}(\a_k)$ and $[\tilde{\a}_{k,r}]_{i} = \exp(-j \frac{N_A+1-2i}{2}\pi\cos(\theta_k)\tilde{\theta}_{k,r})$. The matrix $\Psib_k$ can then be simplified as $\Psib_k = \D_k\B_k \D_k^\herm$, where matrix $\B_k = \mathbb{E}\left(\tilde{\A}_{k,r}\D_k^\herm \R_\x \D_k \tilde{\A}_{k,r}^\herm\right)$. Let $\C_k = \D_k^\herm \R_x \D_k$, then the entries of $\B_k$ can be given as
\begin{align}
   [\B_k]_{i,j} &=  \mathbb{E}\left(\sum_{l}\sum_k[\tilde{\A}_{k,r}]_{i,l}[\C_k]_{l,k}[\tilde{\A}_{k,r}^\herm]_{k,j}\right) \notag, \\
   & = \sum_{l}\sum_k[\C_k]_{l,k}\mathbb{E}\left([\tilde{\A}_{k,r}]_{i,l}[\tilde{\A}_{k,r}^*]_{j,k}\right)\notag,\\
   & = \sum_l\sum_k[\C_k]_{l,k}[\Z_k^{i,j}]_{l,k}\notag,\\
   & = \a_k^\herm\left(\R_\x \odot \Z_k^{i,j}\right)\a_k \label{eq:B_k_i_j},
\end{align}
where $[\Z_{k}^{i,j}]_{l,k} = \psi_{\tilde{\theta}_{k,r}}(((i-j)-(l-k))\pi \cos(\theta_k))$ is the characteristic function matrix of random variable $\tilde{\theta}_{k,r}$. The matrix $\Z_k^{i,j}$ has a Toeplitz structure and hence can be computed easily by only computing $2N_A-1$ entries rather than $N_A^2$. 

The subspace-based algorithms \cite{WeightedSubspace,parametricLocalization,CMS} are based on the fact that the signal space of matrices $\Psib_k \forall k$ is orthogonal to the noise space of the receive data covariance matrix, i.e., 
\begin{align}
    \R_\y = \E_s\Lambdab_s\E_s^\herm + \E_n\Lambdab_n\E_n^\herm,
\end{align}
where $\E_s = \begin{bmatrix}
    \e_1 & \e_2 & \cdots & \e_q
\end{bmatrix}$ represents the signal space and $\E_n = \begin{bmatrix}
    \e_{q+1} & \cdots & \e_{N_A}
\end{bmatrix}$ represents the noise space. The matrices $\Lambdab_s$ and $\Lambdab_n$ are diagonal matrices with corresponding eigenvalues of signal and noise space, respectively. The spectrum can be plotted through
\begin{align}
    P(\theta,\sigma_\theta) = \norm{\Gammab\Psib(\theta,\sigma_\theta)}_\mathrm{F}^2 \label{eq:Spectrum}.
\end{align}
For DISPARE algorithm \cite{CMS}, $\Gammab = \hat{\E}_n$ while for Zoubir's method, $\Gammab =\hat{\R}^{-1}_\Y $, where hat represents the estimated matrices from the received data. In addition, in DSPE \cite{parametricLocalization}, $\Psib^{1/2}(\theta,\sigma_\theta)$ is used as a signal space component in evaluating \eqref{eq:Spectrum}. Nevertheless, subspace models are prone to high \gls{RMSE} at low \gls{SNR} and depend on the small-spread approximation. Moreover, the signal space dimension threshold $q$ is difficult to estimate as it is directly dependent on the target spread. In the following section, we will define \gls{MST} based array response vector modeling, which will remove the requirement of small spread approximation. 

}
\section{\Gls{MST} based data model}
{
In \gls{MST}, the objective is to transform the nonlinear directional phase component present inside the array response vectors into a linear phase component. This can be done by truncating the Jacobi angle-based expansion of the entries of array response vectors, i.e., $e^{\iota z\sin(\theta)} = \sum_{n=-\infty}^\infty J_n(z)e^{\iota n \theta}$. For all the elements of the array response vector, this can be written as a matrix vector multiplication \cite{MST_Base}:
\begin{align}
    \a_k = \G\v_k + \varepsilon,
\end{align}
where $\G \in \mathbb{C}^{N_A \times 2Q+1}$ is the sampling matrix which is entirely dependent on the geometry of the antenna array and is independent of the target direction, and hence can be computed offline. $\v_k \in \mathbb{C}^{2Q+1}$ is Vandermonde structure vector defined as $\v_k = \begin{bmatrix}
    e^{-jQ \theta_k}, \cdots , e^{jQ \theta_k}
\end{bmatrix}$, and $\varepsilon$ is the modeling error originating from the consequence of the truncation of the Jacobi Anger expansion. As the mode number $2Q+1 \to \infty$, $||\varepsilon|| \to 0.$ This transformation from a nonlinear sine function to a linear phase allows us to use an FFT-based algorithm to get the coarse estimates of the parameters, which will be elaborated in the next section. Assuming the mode number is large enough such that the modeling error $\varepsilon$ can be safely ignored, the matrix $\Psib_k$ can be written as $\Psib_k = \G\D_k\B_k\D_k^\herm\G^\herm$, where $\D_k = \mathrm{diag}\left(\v_k\right),$ $[\B_k]_{i,j} = \v_k^\herm\left(\G^\herm\R_\x\G \odot \Z_k^{i,j}\right) \v_k$, where now the matrix $\Z_k^{i,j}$ is $2Q+1 \times 2Q+1$ matrix, and its entries are given by $[\Z_k^{i,j}]_{m,n} = \psi_{\tilde{\theta}_{k,r}}(((i-j)-(n-m))$. Note that now, the matrix $\Z_k^{i,j}$ is completely independent of the mean direction $\theta_k$ and only relies on the angular spread $\sigma_{\theta_k}$. 
}
\section{Parameter Estimation}
{
    In our setting, we are interested in the estimation of angular parameters of targets present in the scene based on the received data covariance matrix: we define the estimation vector as $\etav = \begin{bmatrix}
        \boldsymbol{\theta}^\tran& \boldsymbol{\sigma}_\theta^\tran
    \end{bmatrix}^\tran \in \mathbb{R}^{2K}$, where $\thetav = \begin{bmatrix}
        \theta_1 & \theta_2 & \cdots & \theta_K
    \end{bmatrix}^\tran$, and $\sigmav_\theta = \begin{bmatrix}
        \sigma_{\theta_1}&\sigma_{\theta_2}&\cdots & \sigma_{\theta_K}
    \end{bmatrix}^\tran$. 
\subsection{Coarse Estimate Stage}
Using the \gls{MST} modeling, the spectrum in \eqref{eq:Spectrum} can be expanded as
\begin{align}
    P(\theta,\sigma_\theta) &= \mathrm{Tr}\left(\Psib(\theta,\sigma_\theta)\Gammab^2 \Psib(\theta,\sigma_\theta)\right)\notag,\\
    = \int_\phi\int_\varphi&\rho(\phi;\theta,\sigma_\theta)\rho(\varphi; \theta,\sigma_\theta)\upsilon_{1}^{\phi,\phi}\upsilon_2^{\phi,\varphi}\upsilon_1^{\varphi,\varphi}\upsilon_3^{\varphi,\phi} d\phi d\varphi, \label{eq:P(theta,sigma_theta)_orig}
\end{align}
where $\upsilon_1^{\phi,\phi} = \v^\herm(\phi)\G^\herm \R_\x\G\v(\phi)$, $\upsilon_2^{\phi,\varphi} = \v^\herm(\phi)\G^\herm\Gammab\G\v(\varphi)$, and $\upsilon_3^{\varphi,\phi} = \v^\herm(\varphi)\G^\herm\G\v(\phi)$. The term $\upsilon_1^{\phi,\phi}$ evaluates the transmitted power towards the look direction $\phi$. Expanding $\upsilon_1^{\phi,\phi}$ yields:
\begin{align}
    \upsilon_1^{\phi,\phi} &= \sum_{q_1 = 1}^{2Q+1}\sum_{q_2=1}^{2Q+1}e^{-j(q_1-1)\phi}[\G^\herm\R_\x\G]_{q_1,q_2}e^{j(q_2-1)\phi}.
\end{align}

Since the angular variables in the exponential terms are identical, they combine as $e^{-j(q_1-1)\phi}e^{j(q_2-1)\phi} = e^{-j(q_1-q_2)\phi}$. By applying the change of variables $d = q_1 - q_2$, where $d \in [-2Q, 2Q]$ represents the diagonal index of the matrix $\J = \G^\herm\R_\x\G$, the double summation collapses into a single sum over the diagonals:
\begin{align}
    \upsilon_1^{\phi,\phi} &= \sum_{d=-2Q}^{2Q} \left( \sum_{q_1-q_2=d} [\J]_{q_1,q_2} \right) e^{-jd\phi} \notag\\
    &= \sum_{d=-2Q}^{2Q} w_d e^{-jd\phi}, 
\end{align}
where $w_d$ is the scalar sum of all elements on the $d$-th diagonal of $\J$. Evaluating this continuous function over a discrete angular grid of $N$ points ($\phi_k = \frac{2\pi k}{N}$) perfectly maps to the strict definition of a 1-Dimensional Discrete Fourier Transform (DFT). Thus, $\upsilon_1^{\phi,\phi}$ can be computed instantaneously across the entire grid via a 1D Fast Fourier Transform (FFT) of the zero-padded vector $\mathbf{w}$.

Conversely, the term $\upsilon_2^{\phi,\varphi}$ contains two distinct angular variables, representing the spatial cross-coupling between directions $\phi$ and $\varphi$ through the weighting matrix $\Gammab$. Letting $\B = \G^\herm\Gammab\G$, we obtain:
\begin{align}
    \upsilon_2^{\phi,\varphi} &= \nu\sum_{q_1=1}^{2Q+1}\sum_{q_2=1}^{2Q+1} e^{-j(q_1-1)\phi}[\B]_{q_1,q_2}e^{j(q_2-1)\varphi},
\end{align}
where $\nu = e^{\iota Q(\phi - \varphi)}$. This structure is mathematically equivalent to applying a forward Fourier transform to the columns of $\B$, followed by an inverse Fourier transform to the resulting rows. To compute this over an $N\times N$ grid, we construct a 2D lookup table $\mathbf{\Upsilon}_2 \in \mathbb{C}^{N \times N}$ defined as
\begin{align}
    \mathbf{\Upsilon}_2 &= N \mathcal{F}^{-T} \Big\{ \mathcal{F}^T \Big\{ \G^\herm\Gammab\G \Big\} \Big\},
\end{align}
where $\mathcal{F}^T\{\cdot\}$ denotes a column-wise forward ]gls{FFT} followed by a matrix transpose, and $\mathcal{F}^{-T}\{\cdot\}$ denotes a column-wise \gls{IFFT} followed by a transpose. Similarly, for the spatial manifold term $\upsilon_3^{\varphi,\phi}$, we define the lookup table $\Upsilonb_3 = N \mathcal{F}^{-1}\Bigl\{\mathcal{F}^T\Bigl\{\G^\herm\G\Bigr\}\Bigr\}$. 

Letting $\Upsilonb_1 = \mathcal{F}\left\{\w\right\}\mathcal{F}^\tran\left\{\w\right\}$, the complete $N \times N$ discrete evaluation matrix can be constructed via the Schur-Hadamard product:
\begin{align}
    \Upsilonb &= {\Re}\Bigl\{\Upsilonb_1 \odot \Upsilonb_2 \odot \Upsilonb_3\Bigr\}. \label{eq:Upsilon}
\end{align}

Assuming uniformly distributed targets $\rho \sim \mathcal{U}[\theta-\sigma_\theta,\theta+\sigma_\theta]$, we divide the spatial region into $L$ equal subintervals with spacing $\Delta = \frac{2\sigma_\theta}{L}$. Exploiting the precomputed lookup table defined in \eqref{eq:Upsilon}, the integral in \eqref{eq:P(theta,sigma_theta)_orig} is elegantly approximated by:
\begin{align}
    P(\theta,\sigma_\theta) \approx \frac{1}{(L+1)^2}\sum_{n_1 = -L/2}^{L/2}\sum_{n_2 = -L/2}^{L/2}[\Upsilonb]_{\frac{\theta}{\Delta}+n_1,\frac{\theta}{\Delta}+n_2}. \label{eq:P(theta,sigma_theta)_approx}
\end{align}
This operation is equivalent to passing a 2D moving-average box filter over the matrix $\Upsilonb$. By executing the entire coarse estimation stage in the discrete Fourier domain, the computational bottleneck of executing $O(N_A^3)$ matrix multiplications for every point in the $(\theta,\sigma_\theta)$ search grid is entirely circumvented. For alternative distributions (e.g., Gaussian), the core matrix $\Upsilonb$ remains identical; only the moving average weighting kernel is modified \cite{MST_Base}.

\subsection{Fine Estimation}
To improve the accuracy of the coarse estimates, we employ a weighted least-squares (WLS) framework \cite{WeightedSubspace}. Assuming the base station possesses prior knowledge of the noise floor $\sigma_b^2$, the unknown parameter vector is defined as $\tilde{\etav} = [\etav^\tran, q_1^2, q_2^2, \cdots, q_K^2]^\tran \in \mathbb{R}^{3K}$. The WLS objective function \cite{WeightedSubspace} to minimize is given by:
\begin{align}
    f(\tilde{\etav}) = \mathrm{Tr}\left(\left(\R\left(\tilde{\etav}\right)\hat{\R}_\y^{-1} - \I \right)^2\right), \label{eq:ObjectiveFunc_finetuning}
\end{align}
where $\R(\tilde{\etav}) = \sum_k q_k^2\Psib_k + \sigma_b^2\I$. Because $f(\tilde{\etav})$ is highly non-linear with respect to both $\theta$ and $\sigma_\theta$, we apply a first-order Taylor series expansion of the theoretical covariance matrix around the initial coarse estimates $\tilde{\etav}_0$:
\begin{align}
    \R\left(\tilde{\etav}_0+\Delta \tilde{\etav}\right) \approx \R(\tilde{\etav}_0) + \sum_{i=1}^{3K}\Delta [\tilde{\etav}]_i\frac{\partial\R}{\partial [\tilde{\etav}]_i}\Big\rvert_{\tilde{\etav}_0}. \label{eq:Taylor}
\end{align}
The initial estimates of the signal strengths $q_k^2$ are provided by the Generalized Capon estimator \cite{GeneralizedCaponEstimator}:
\begin{align}
    \hat{q}_k^2 = \frac{1}{\mathrm{eig}_\mathrm{max}\left(\hat{\R}_\y^{-1}\hat{\Psib}_k\right)}.
\end{align}
Substituting \eqref{eq:Taylor} into \eqref{eq:ObjectiveFunc_finetuning} transforms the objective into a locally quadratic function with respect to the step vector $\Delta \tilde{\etav}$. Taking the derivative with respect to each parameter perturbation $[\Delta \tilde{\etav}]_k$ and equating to zero yields:
\begin{align}
    \frac{\partial f}{\partial [\Delta \tilde{\etav}]_k} &= 2\mathrm{Tr}\left(\left[\R(\tilde{\etav}_0+\Delta\tilde{\etav})\hat{\R}_\y^{-1} - \I\right]\frac{\partial\R}{\partial [\tilde{\etav}]_k}\Bigr|_{\tilde{\etav}_0}\hat{\R}_\y^{-1}\right)=0.
\end{align}
Isolating the summation terms containing $\Delta \tilde{\etav}_i$, we obtain a robust system of linear equations mapping directly to the Gauss-Newton update step:
\begin{align}
    \sum_{i=1}^{3K}\Delta [\tilde{\etav}]_i\mathrm{Tr}&\left( \frac{\partial\R}{\partial[\tilde{\etav}]_i}\hat{\R}_\y^{-1}\frac{\partial\R}{\partial[\tilde{\etav}]_k}\hat{\R}_\y^{-1}\right) \notag\\
    &= \mathrm{Tr}\left(\left[\I - \R(\tilde{\etav}_0)\hat{\R}_\y^{-1}\right]\frac{\partial\R}{\partial [\tilde{\etav}]_k}\hat{\R}_\y^{-1}\right). \label{eq:GaussNewtonExpansion}
\end{align}
Expressed in compact matrix notation, \eqref{eq:GaussNewtonExpansion} reduces to:
\begin{align}
    \H(\tilde{\etav}_0)\Delta \tilde{\etav} &= \r \quad \Longrightarrow \quad \Delta \tilde{\etav} = \H^{-1}\left(\tilde{\etav}_0\right)\r,
\end{align}
where the elements of the approximate Hessian $\H \in \mathbb{R}^{3K \times 3K}$ and the gradient vector $\r \in \mathbb{R}^{3K}$ are respectively defined as:
\begin{align}
    [\H(\tilde{\etav}_0)]_{i,k} &= \mathrm{Tr}\left(\frac{\partial\R}{\partial[\tilde{\etav}]_k}\Bigr\rvert_{\tilde{\etav}_0}\hat{\R}_\y^{-1}\frac{\partial\R}{\partial [\tilde{\etav}]_i}\Bigr|_{\tilde{\etav}_0}\hat{\R}_\y^{-1}\right), \notag\\
    \r_i &= \mathrm{Tr}\left(\left[\I - \R(\tilde{\etav}_0)\hat{\R}_\y^{-1}\right]\frac{\partial\R}{\partial [\tilde{\etav}]_i}\Bigr\rvert_{\tilde{\etav}_0}\hat{\R}_\y^{-1}\right).
\end{align}
The parameter vector $\tilde{\etav}_0$ is then updated as
\begin{align}
    \tilde{\etav}_0 \leftarrow \tilde{\etav}_0  +\zeta \Delta \tilde{\etav},
\end{align}
where $\zeta$ is the learning rate of Gauss-Newton iteration. The closed-form analytical derivatives of the MST covariance matrix are provided in the Appendix. Both Coarse and Fine estimation algorithms are summarized in Algorithm \eqref{alg:proposed_estimator}.
\subsection{Computational Complexity Analysis}
The computational complexity of the proposed two-stage estimator is analyzed and compared against three classical baseline estimators: the \textit{Generalized Capon (GC)} \cite{GeneralizedCaponEstimator}, \textit{DISPARE} \cite{CMS}, and \textit{Zoubir's} \cite{Zoubir} method. Let $N_\theta$ and $N_\sigma$ denote the number of search points for the nominal DOA and angular spread grids, respectively.

In the conventional 2D grid-search estimators (GC, DISPARE, and Zoubir's method), the theoretical spatial covariance matrix $\Psib(\theta, \sigma_\theta)$ must be explicitly reconstructed for every point in the search space. Furthermore, evaluating the respective spatial spectra at each point demands computationally intensive matrix operations. For instance, Zoubir's method requires computing the Frobenius norm of $\hat{\R}_\y^{-1}\Psib$ \cite{Zoubir}, DISPARE requires projecting $\Psib$ onto the noise subspace, and the standard GC method requires computing the principal eigenvalue of $\hat{\R}_\y^{-1}\Psib$. All three methodologies incur an asymptotic complexity of $\mathcal{O}(N_A^3)$ per grid point, resulting in a total computational burden of $\mathcal{O}(N_\theta N_\sigma N_A^3)$. Because the grid resolution is multiplicatively coupled to the array dimension, these legacy estimators become prohibitively expensive for massive MIMO arrays with large $N_A$.

The proposed estimator radically reduces the computational burden by decoupling the 2D search grid from the array dimensions. In the coarse estimation stage, mapping the spatial manifold to the Fourier domain via MST requires 1D and 2D FFT operations, bounded by $\mathcal{O}(N^2 \log N)$, where $N$ is the FFT grid dimension. Once the master lookup table $\Upsilonb$ is constructed, calculating the spatial spectrum $P(\theta, \sigma_\theta)$ using the moving average requires only four scalar additions, yielding an algorithmic complexity of $\mathcal{O}(1)$ per search point\cite{Application_MST}. 

In the subsequent fine-estimation stage, the Gauss-Newton optimization entirely circumvents the need for dense grid evaluations. Constructing the approximate Hessian $\H$ and the gradient vector $\r$ requires computing analytical matrix derivatives and traces, which take approximately $\mathcal{O}(K N_A^3)$ operations per iteration. Assuming the algorithm converges in $N_{iter}$ iterations, the total complexity of the fine stage is bounded by $\mathcal{O}(N_{iter} K N_A^3)$. 

Overall, the proposed estimator operates with a complexity of $\mathcal{O}(N^2 \log N + N_\theta N_\sigma + N_{iter} K N_A^3)$. By transforming the multiplicative coupling of the baseline algorithms into an additive relationship, the proposed method guarantees superior computational tractability and scalability, particularly for modern systems employing large-scale antenna arrays.

\begin{algorithm}[t]
\caption{Proposed MST-Based Coarse-to-Fine Parameter Estimation}
\label{alg:proposed_estimator}
\begin{algorithmic}[1]
\REQUIRE $\hat{\R}_\y$, $\G$, $\Gammab$, $\R_\x$, $\sigma_b^2$, $N$, $K$, learning rate $\zeta$.
\STATE \textbf{Phase 1: 2D FFT Coarse Estimation}
\STATE Compute diagonals sum vector $\w$ from $\A = \G^\herm\R_\x\G$.
\STATE Compute $\Upsilonb_1 = \mathcal{F}\{\w\}\mathcal{F}^\tran\{\w\}$ via 1D FFT.
\STATE Compute $\Upsilonb_2 = N \mathcal{F}^{-T} \{ \mathcal{F}^T \{ \G^\herm\Gammab\G \} \}$ via 2D FFT.
\STATE Compute $\Upsilonb_3 = N \mathcal{F}^{-1} \{ \mathcal{F}^T \{ \G^\herm\G \} \}$ via 2D FFT.
\STATE Construct the spatial matrix $\Upsilonb = \Re\{\Upsilonb_1 \odot \Upsilonb_2 \odot \Upsilonb_3\}$.
\FOR{each $\sigma_\theta$}
    \STATE Determine window length $L = \mathrm{round}(2\sigma_\theta / \Delta)$.
    \STATE Evaluate spectrum $P(\theta, \sigma_\theta)$ by passing an $L \times L$ moving average box filter along the main diagonal of $\Upsilonb$ using \eqref{eq:P(theta,sigma_theta)_approx}.
\ENDFOR
\STATE Identify the $K$ dominant peaks to obtain initial estimates $\etav_0 = [\hat{\theta}_1, \dots, \hat{\theta}_K, \hat{\sigma}_1, \dots, \hat{\sigma}_K]^\tran$.

\STATE \textbf{Phase 2: Gauss-Newton Fine Tuning}
\STATE Initialize signal powers $\hat{q}_k^2 = 1 / \mathrm{eig}_\mathrm{max}\left(\hat{\R}_\y^{-1}\hat{\Psib}_k(\etav_0)\right)$.
\STATE Form the initial parameter vector $\tilde{\etav} = [\etav_0^\tran, \hat{q}_1^2, \dots, \hat{q}_K^2]^\tran$.
\WHILE{convergence criteria not met \AND iterations $<$ max}
    \STATE Compute analytical derivatives $\frac{\partial\R}{\partial[\tilde{\etav}]_i}$ for $i \in \{1, \dots, 3K\}$.
    \STATE Construct approximate Hessian $\H(\tilde{\etav})$ and gradient $\r(\tilde{\etav})$.
    \STATE Compute step vector $\Delta\tilde{\etav} = \H^{-1}\r$.
    \STATE Update estimates: $\tilde{\etav}_0 \leftarrow \tilde{\etav}_0 + \zeta\Delta\tilde{\etav}$.
\ENDWHILE
\ENSURE Fine estimates of DOAs, angular spreads, and signal powers $\tilde{\etav}$.
\end{algorithmic}
\end{algorithm}

}

    \section{Optimization Problem and its solution}
    {
    From both coarse and fine tuning algorithms, it is evident that the estimator performance is inherently dependent on the transmit beamforming through the transmit covariance matrix $\R_\x = \V\V^\herm$.
    In our assumed \gls{ISAC} settings based on \gls{FD} \gls{BS}, the designed beamformer must be able to satisfy 1) users' data rate requirements, 2) \gls{SI} power threshold at the \gls{RX} RF chains after the analog cancellation to avoid RF chain saturation with \gls{SI} power, and 3) it must illuminate all of the targets' extent, which depends on their angular spreads. For target illumination, we define the objective function to minimize the weighted distance between our designed beamformer $\V$ and the optimal beamformer $\V_\mathrm{ideal}$ (which illuminates all the targets): $g(\V) = \mathbb{E}\left(\norm{\H_{\mathrm{Rad},p}\left(\V_{\mathrm{ideal}} - \V\right)}_\mathrm{F}^2\right)$. To find $\V_{\mathrm{ideal}}$, we define $\R_{{\mathrm{Rad}}} = \mathbb{E}\left(\H_{\mathrm{Rad},p}^\herm\H_{\mathrm{Rad},p}\right)$. Utilizing the estimates of $\etav$ from the previous timeslot, we can construct the estimated $\R_\mathrm{Rad}$ as 
    \begin{align}
        \hat{\R}_\mathrm{Rad} = \sum_{k}\G\hat{\D}_k\hat{\tilde{\B}}_k\hat{\D}_k^\herm\G^\herm,
    \end{align}
    where $[\hat{\tilde{\B}}_k]_{m,n} = \psi_{\hat{\tilde{\theta}}_{k,r}}(m-n)$. Note that the matrix $\hat{\tilde{\B}}_k$ is different from $\B_k$ in the sense that here, each entry is independent of $\R_\x$. The eigen decomposition of $\hat{\R}_{\mathrm{Rad}}$ will yield $\R_\mathrm{Rad}  = \Q_\mathrm{Rad}\Lambdab_\mathrm{Rad}\Q^\herm_{\mathrm{Rad}}$. The optimal beamformer for target illumination will then be $\V_\mathrm{ideal} = \Q_\mathrm{Rad}$. The objective function $g(\V)$ can then be rewritten as
    \begin{align}
        g(\V) = \norm{\Lambdab_{\mathrm{Rad}}^{1/2}\left(\I - \Q_{\mathrm{Rad}}^\herm\V\right)}_\mathrm{F}^2.
    \end{align}
     To incorporate the data rate as well as \gls{SI} constraints, we define the following optimization problem
\begin{subequations}\label{opt_prob_1}
    \begin{align}
    \min_{\V} \quad & \quad g(\V) \label{Opt_prob_1:objective}\\
          \textrm{s.t.} \quad & \Gamma_u = \log_2\left(1+\gamma_u^\mathrm{sinr}\right) \geq \gamma_u,\quad \forall u \in \mathcal{L}, \label{opt_prob_1:orig_rate_constraint}\\
           &\quad  \norm{\left(\H_{\mathrm{SI}} + \C\right)\V}_{2,\infty}  \leq \sqrt{\lambda_\mathrm{SI}}\label{opt_prob_1:SI_consraint}\\
&\quad\left\lvert\left\lvert\V\right\rvert\right\rvert_\rm{F}^2 \leq \rm{P}_b, \label{opt_prob_1:Power_constraint}
    \end{align}
\end{subequations}
 where constraint \eqref{opt_prob_1:orig_rate_constraint} represents the data rate constraints and $\gamma_u^\mathrm{sinr} = \frac{\norm{\h_u^\herm\v_u}_2^2}{\sum_{u' \neq u}^{N_A} \norm{\h_u^\herm\v_{u'}}_2^2 + \sigma_u^2}$ is the $u$-the user data rate requirement, \eqref{opt_prob_1:SI_consraint} limits the \gls{SI} power at the receiver RF chain after analog cancellation to avoid RF signal saturation, and $\lambda_\mathrm{SI}$ denotes the \gls{SI} saturation threshold. Here $\norm{.}_{2,\infty}$ is the induced matrix norm given as $\norm{\A}_{2,\infty} = \max_{1\leq i\leq m}\norm{[\A]_{i,:}}_2$ for matrix $\A \in \mathbb{C}^{m\times n}$. The constraint \eqref{opt_prob_1:Power_constraint} represents the total transmit power constraint. 

    The objective function \eqref{Opt_prob_1:objective} and \gls{SI} power constraint \eqref{opt_prob_1:SI_consraint} are convex in $\V$. To expose the convexity of the data rate constraint \eqref{opt_prob_1:orig_rate_constraint}, we follow the strategy described in \cite{SINR_Reformulation1}. \eqref{opt_prob_1:orig_rate_constraint} can be written as follows $   \abs{\h_u^\herm\v_u}^2 \geq(2^{\gamma_u} - 1) \bigl(\sum_{u'\neq u }^{N_A}\abs{\h^\herm_u\v_u'}^2 +\sigma_u^2\bigr)$.
    This inequality is non-convex in general; however, the phase of the term $\h_u^\herm\v_u$ does not affect the optimal value of the constraint \cite{SINR_Reformulation1}, hence, we can only focus on the real part of $\h_u^\herm \v_u$, i.e., $\Re\{\h_u^\herm\v_u^\herm\}$. Define the matrix $\tilde{\V}_u = \V\left(\I - \diag(\e_u)\right)$, where $\e_u$ is the $u^\mathrm{th}$ column of identity matrix. Then the rate constraint can be reformulated as \gls{SOC}
    \begin{align}
        \begin{bmatrix}
            \tilde{\V}_u^\herm \h_u \\\sigma_u \\ \sqrt{\frac{1}{2^{\gamma_u} - 1}} \Re\left\{\h_u^\herm \v_u\right\}
        \end{bmatrix} &\succeq_{\mathrm{SOC}} 0,  \quad \forall \,u \in \mathcal{L}. \label{eq:Rate_Constraint}
    \end{align}

    Replacing original rate constraint \eqref{opt_prob_1:orig_rate_constraint} with \eqref{eq:Rate_Constraint} will make the original problem \eqref{opt_prob_1} completely convex in $\V$ and can be easily solved through optimization tools such as CVX.
        
}

\section{Simulation Results}
{
In this section, we present the simulation results for the proposed estimator and the designed beampattern obtained via the optimization framework defined in \eqref{opt_prob_1}. We choose RF saturation level $\lambda_\mathrm{SI}$ to be $30\,$dB higher than the noise floor $\sigma_b^2 = -90\,$dBm, $Q = 30$, and $N = 2048$ such that the coarse resolution of $\theta$ is $360/N = 0.17^\circ$. In Fig. \ref{fig:TX_beampattern}, we present the designed beampattern by solving \eqref{opt_prob_1}, where we choose $\gamma_u = 4\,$bps/Hz. The users' directions are $-40^\circ, -20^\circ$, and $-10^\circ$, while the target direction is $\theta_1 = 20^\circ$ with $\sigma_{\theta_1} = 8^\circ$. The RF saturation threshold is set to $30\,$dB above the noise floor \cite{ISACATIQ}. The proposed optimization framework can design an adaptive beamformer that yields a beampattern nearly identical to the ideal one while meeting data-rate and RF-saturation constraints. 

\begin{figure}[t]
    \centering
    \includegraphics[width=1\linewidth]{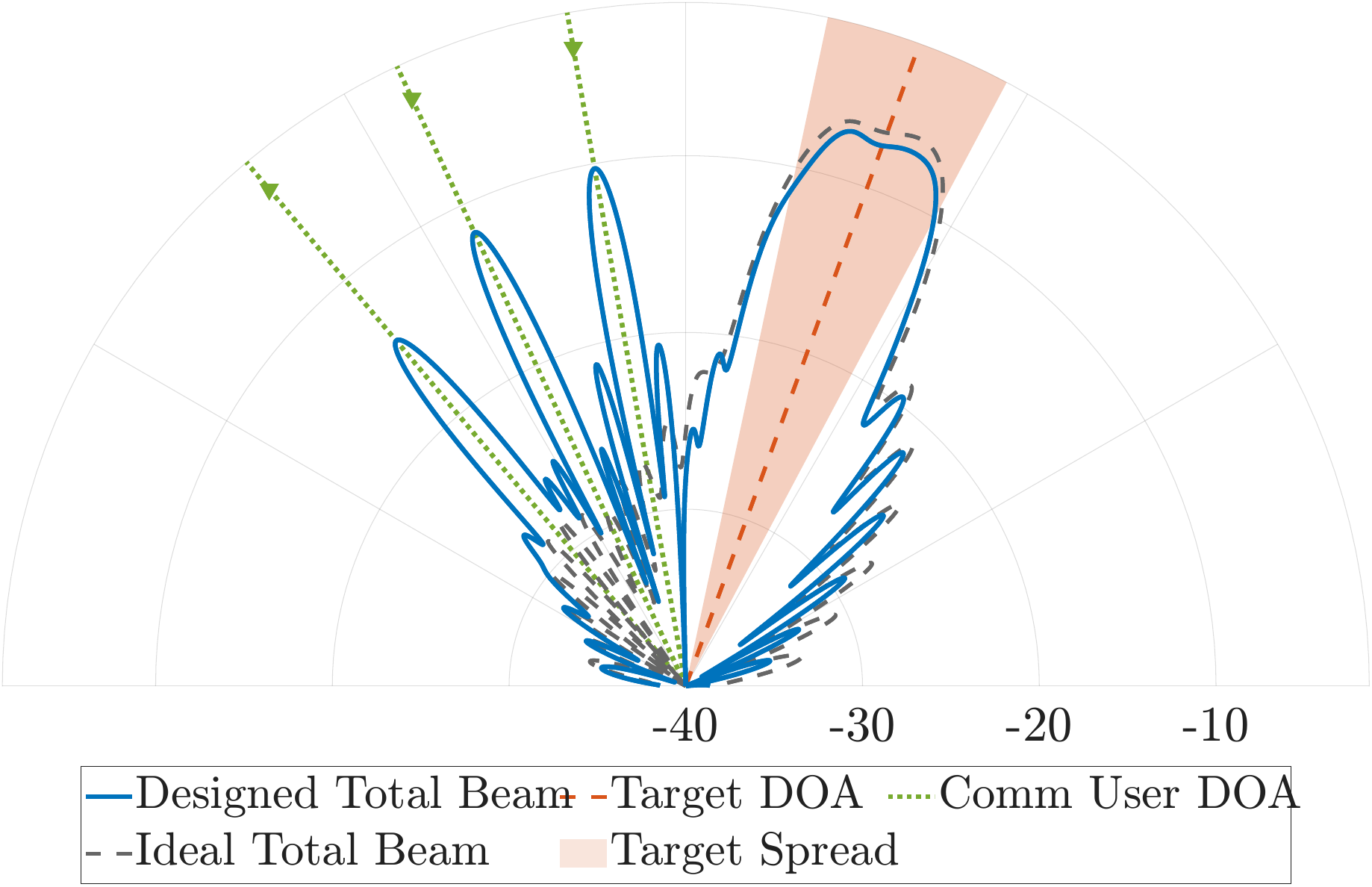}
    \caption{\gls{TX} beampattern from proposed optimization vs beampattern generated through $\V_{\mathrm{ideal}}$. We choose $N_T = 32$, with $\gamma_u = 0.5$bps/Hz $\forall u$, \gls{SNR} of $25\,$dB. }
    \label{fig:TX_beampattern}
\end{figure}

In all the following results, we considered data captured over a single \gls{OFDM} slot to estimate the covariance matrix. Moreover, we choose $N_A = 16$ antennas, and $\zeta = 0.01$ for the fine-tuning algorithm. For comparison, we compare the performance of both coarse estimator (denoted by \textit{Coarse FFT}) and the complete algorithm presented in \eqref{alg:proposed_estimator} (denoted by \textit{Proposed}) against three baseline schemes: 1) \textit{Zoubir} method \cite{Zoubir}, 2) \textit{DISPARE} algorithm \cite{CMS}, and 3) Generalized Capon (\textit{GC}) estimator \cite{GeneralizedCaponEstimator}. Note that all these estimators were not developed for the considered \gls{MIMO} settings in this paper. We have extended these algorithms to the assumed \gls{MIMO} settings to have a fair comparison. For all upcoming results, we ran a Monte Carlo simulation with $300$ runs. The grid size of theta is between $\theta_{\mathrm{min}}-10^\circ$ and $\theta_{\mathrm{max}}+10^\circ$ with step size of $0.1^\circ$. For angular spread, the grid ranges from $0^\circ$ to $10^\circ$ with a step size of $0.1^\circ$.

\begin{figure}[t]
\centering
\begin{subfigure}[t]{1\linewidth}
    \includegraphics[width=\linewidth]{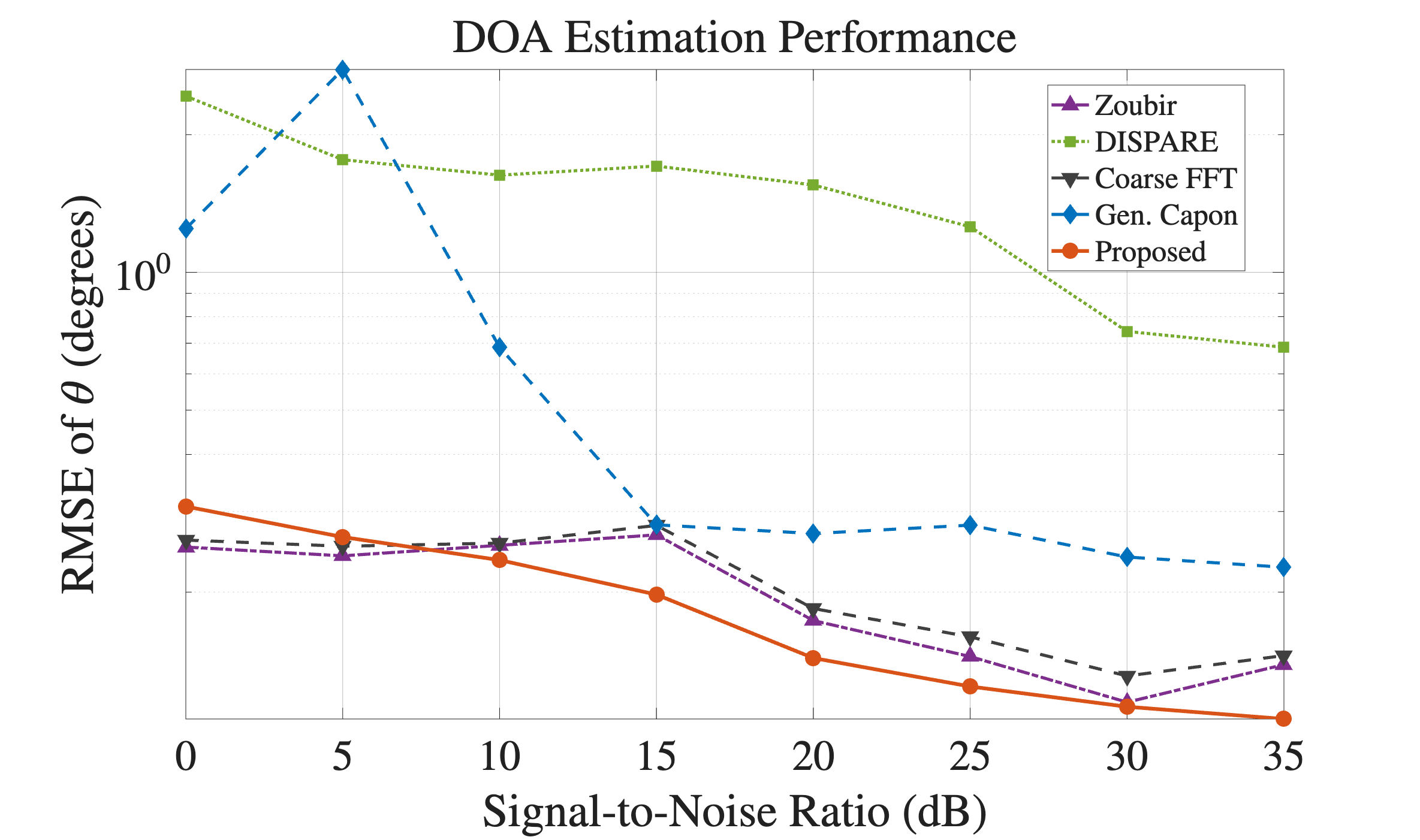}
         \caption{\gls{RMSE}($\theta$) vs \gls{SNR} of the proposed algorithm and different baseline schemes.} 
         \label{fig:RMSE_Theta_vs_SNR}
\end{subfigure}
\hfill
\begin{subfigure}[t]{1\linewidth}
    \includegraphics[width=\linewidth]{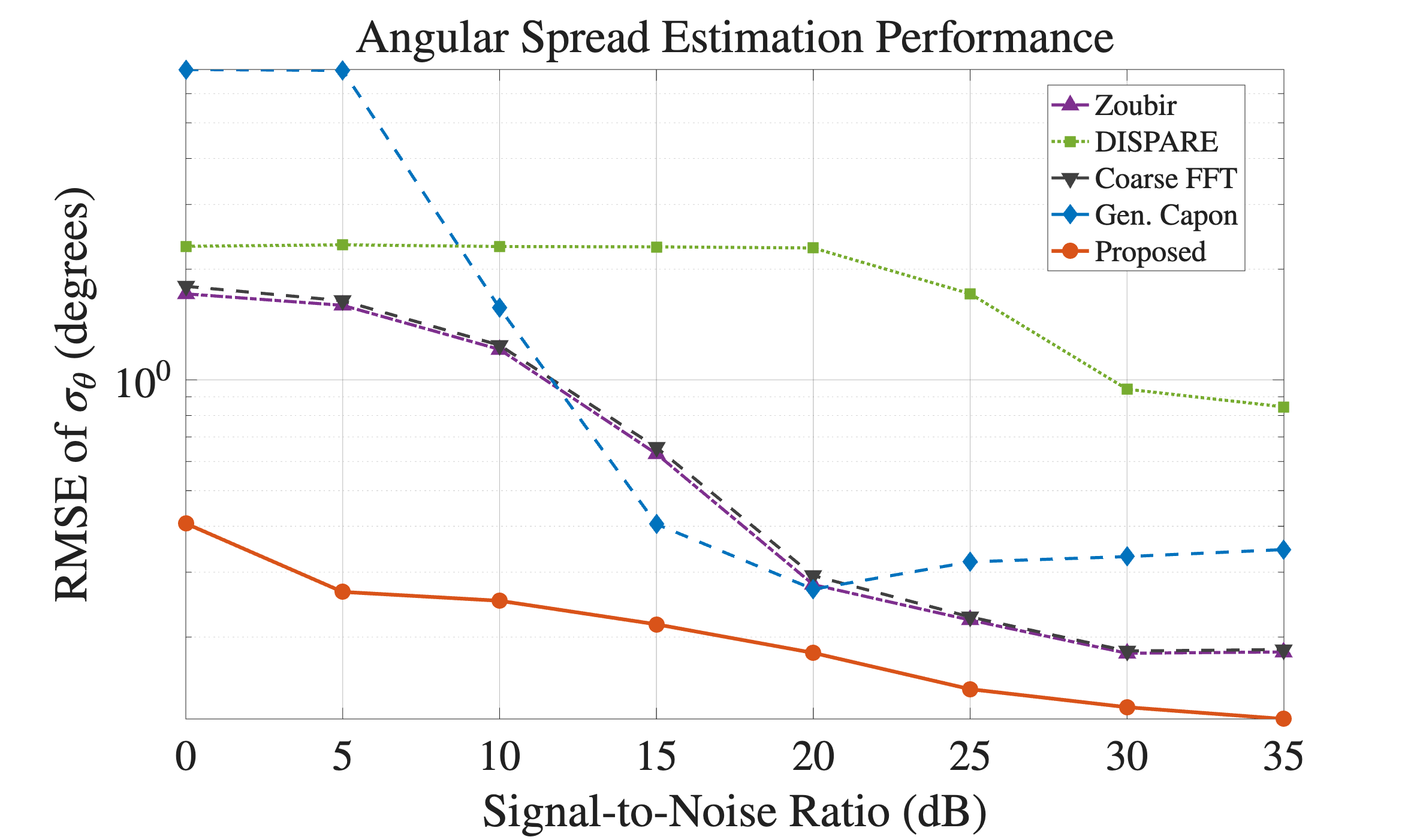}
         \caption{\gls{RMSE}($\sigma_\theta$) vs \gls{SNR} of proposed algorithm compared and different baseline schemes.} 
         \label{fig:RMSE_Sigma_theta_vs_SNR}
\end{subfigure}
\caption{\gls{RMSE} of mean angle and angular spread of proposed algorithm as compared to the baseline subspace-based algorithms. The rate requirement is chosen to be $\gamma_u = 10^{-3}\,$bps/Hz $\forall u$.} 
\label{fig:RMSE_vs_SNR}
\end{figure}

In Fig. \ref{fig:RMSE_vs_SNR}, we compare the performance of the proposed algorithm against different subspace-based baseline schemes. For this result, we considered a single target with $\theta_1 = 20^\circ$ with $\sigma_{\theta_1} = 3^\circ$ and $d_1 = 20\,$m with $\sigma_{d_1} = 4\,$m, with two users located at $-40^\circ$ and $-20^\circ$ with both distances assumed to be $50\,$m  respectively. From the error performance of direction $\theta$, the proposed estimator closely tracks the \textit{Zoubir} and \textit{Coarse FFT} method for all values of \gls{SNR}, with both latter methods providing slightly better error performance in the low \gls{SNR} regime of $\leq 8\,$dB. Moreover, in the low \gls{SNR} regime, fine-tuning did not improve the estimates over the coarse estimates; performance worsened. This is inherently due to the fact that at low \gls{SNR}, the objective function \eqref{eq:ObjectiveFunc_finetuning} is not a smooth function, and hence a small learning rate is necessary to improve the estimator performance. Nevertheless, after \gls{SNR} = $10\,$dB, the proposed estimator outperforms all other estimators. In Fig. \ref{fig:RMSE_Sigma_theta_vs_SNR}, the \gls{RMSE} of angular spread is plotted. For all values of \gls{SNR}, the proposed estimator consistently provides better error performance than all other schemes. For instance, at \gls{SNR} = $0\,$dB, the proposed method achieves nearly three times the performance of the second-best estimator, \textit{Zoubir} (\gls{RMSE} of $0.4^\circ$ compared to $1.1^\circ$). This shows that the fine-tuning procedure yields a greater improvement in accuracy for $\sigma_\theta$ than for $\theta$.

\begin{figure}[t]
\centering
\begin{subfigure}[t]{1\linewidth}
    \includegraphics[width=\linewidth]{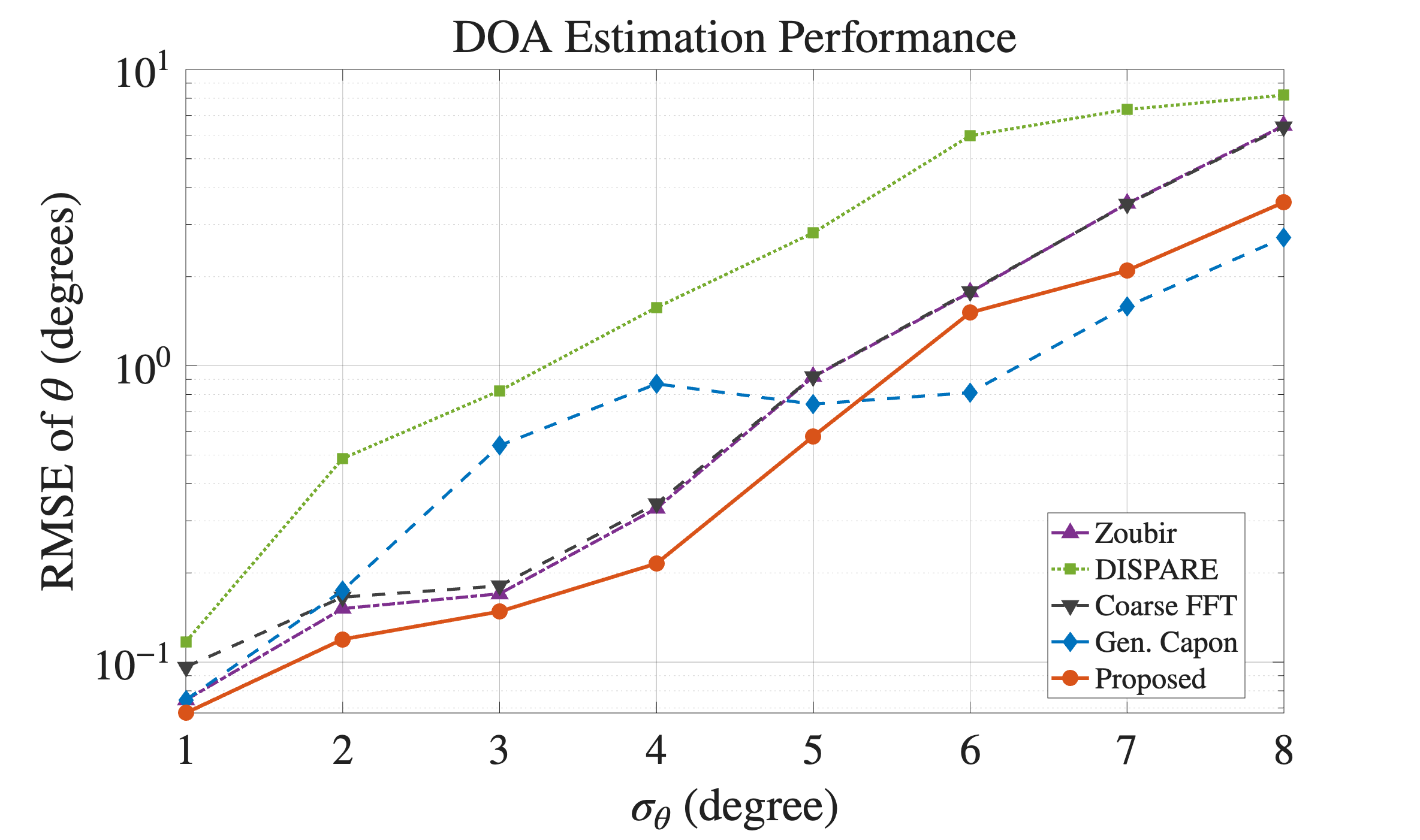}
         \caption{$\frac{1}{K}\sum_k\text{\gls{RMSE}} (\theta_k)$ vs true $\sigma_\theta$.} 
         \label{fig:RMSE_Theta_vs_sigma_theta}
\end{subfigure}
\hfill
\begin{subfigure}[t]{1\linewidth}
    \includegraphics[width=\linewidth]{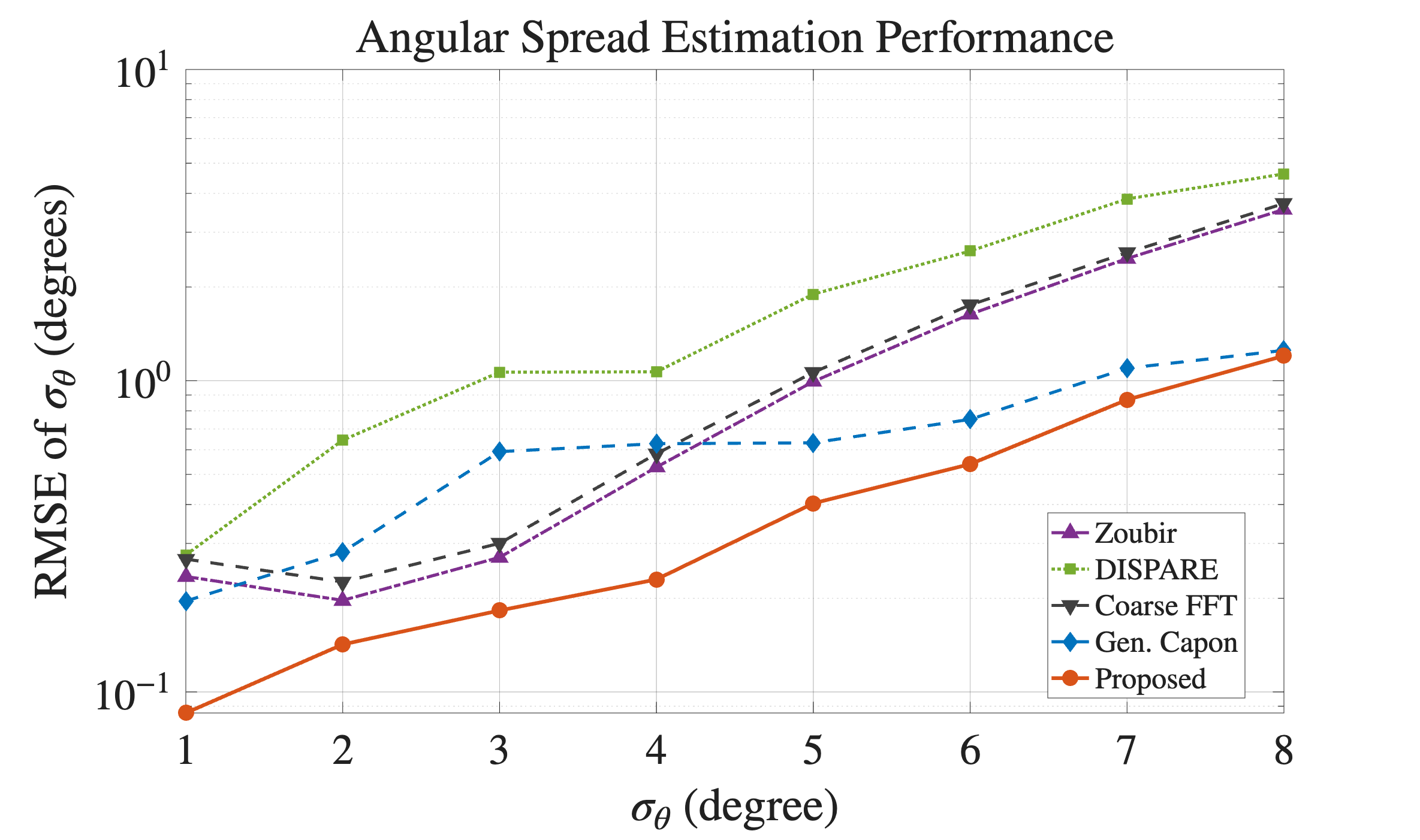}
         \caption{$\frac{1}{K}\sum_k\text{\gls{RMSE}} (\sigma_{\theta_k})$ vs true $\sigma_\theta$.} 
         \label{fig:RMSE_Sigma_theta_vs_sigma_theta}
\end{subfigure}
\caption{\gls{RMSE} of mean angle and angular spread vs true $\sigma_\theta$ of proposed algorithm and comparison with baseline schemes. The rate requirements we choose here are $\gamma_u = 0.5\,$bps/Hz $\forall u$.} 
\label{fig:RMSE_vs_sigma_theta}
\end{figure}

Figure \ref{fig:RMSE_vs_sigma_theta} evaluates the error performance of the estimators as a function of the true angular spread. The simulation considers a two-target scenario localized at $(\theta_1, \theta_2) = (0^\circ, 20^\circ)$, two communication users with minimum rate constraints of $\gamma_u = 0.5$~bps/Hz, and an \gls{SNR} of $25$~dB. The horizontal axis represents the angular spread of both targets.

Figure \ref{fig:RMSE_Theta_vs_sigma_theta} illustrates the Direction of Arrival (DOA) estimation \gls{RMSE}. With the exception of the \textit{GC} method, all estimators exhibit a monotonically increasing error as target dispersion widens. Notably, for $\sigma_\theta \le 5^\circ$, the proposed algorithm yields the highest accuracy across all estimators. However, at extreme spreads ($\sigma_\theta > 5^\circ$), the \textit{GC} method overtakes the proposed estimator. We speculate that this is due to the metric $\frac{1}{\lambda_\mathrm{max}}$ inherent to the \textit{GC} estimator, which introduces bias into the estimate and may favor larger angular-spread targets. 

Figure \ref{fig:RMSE_Sigma_theta_vs_sigma_theta} presents the angular spread estimation error, demonstrating the strict dominance of the proposed method across all simulated conditions. At a narrow spread of $\sigma_\theta = 1^\circ$, the proposed estimator achieves an \gls{RMSE} of approximately $0.1^\circ$, halving the error of the baseline algorithms. While the \textit{GC} method initially degrades in performance, it surpasses both \textit{Zoubir} and the \textit{Coarse \gls{FFT}} algorithms for $\sigma_\theta \geq 4^\circ$. Nevertheless, the proposed fine-tuning stage maintains superior performance across the entire regime and is more resilient to a wide range of angular spreads.
    \section{Conclusion}
    { We consider the \gls{FD}-\gls{ISAC} system capable of estimating angular parameters such as mean angle and spatial spread from distributed clustered targets using communication waveforms. We demonstrate that our proposed methodology successfully outperforms several baseline schemes and provides three times better performance in the low \gls{SNR} regime for spread estimation. We also developed an optimization framework for designing \gls{TX} beamformer to fully illuminate the targets. Our results show that the proposed beampattern is nearly identical to the ideal sensing beampattern even after allocating resources to meet data-rate requirements and minimizing \gls{SI} power. 
    }
}

\appendix[Analytical Derivatives of the Covariance Matrix]
    
The theoretical covariance matrix modeled at the parameter estimates $\tilde{\etav}$ is defined as:
\begin{align}\R(\tilde{\etav}) &= \sum_{k=1}^K q_k^2 \Psib_k(\theta_k, \sigma_{\theta_k}) + \sigma_b^2\I, \label{eq:R_eta_def}\end{align}
The partial derivatives of $\R(\tilde{\etav})$ with respect to the $k$-th target's signal power ($q_k^2$), nominal Direction of Arrival ($\theta_k$), and angular spread ($\sigma_{\theta_k}$) are derived below.
\subsection{Derivative with Respect to Signal Power ($q_k^2$)}
Since the covariance matrix is a linear combination of the spatial signatures weighted by their respective signal powers, the derivative with respect to $q_k^2$ is straightforward:
\begin{align}\frac{\partial \R(\tilde{\etav})}{\partial q_k^2} &= \Psib_k(\theta_k, \sigma_{\theta_k}) = \G \D_k \B_k \D_k^\herm \G^\herm.\end{align}
\subsection{Derivative with Respect to Nominal DOA ($\theta_k$)}
The nominal DOA $\theta_k$ implicitly defines both the virtual steering matrix $\D_k$ and the spatial spreading matrix $\B_k$. Applying the chain rule to the spatial signature $\Psib_k$ yields:
\begin{align}
\frac{\partial \R(\tilde{\etav})}{\partial \theta_k} &= q_k^2 \frac{\partial \Psib_k}{\partial \theta_k} \notag \notag\\
= q_k^2 \G& \Bigl( \dot{\D}_k \B_k \D_k^\herm + \D_k \dot{\B}_k \D_k^\herm + \D_k \B_k \dot{\D}_k^\herm \Bigr) \G^\herm, \label{eq:dPsi_dtheta}
\end{align}
where the overdot notation denotes the partial derivative with respect to $\theta_k$ (i.e., $\dot{(\cdot)} = \frac{\partial (\cdot)}{\partial \theta_k}$). The matrix $\dot{\D}_k$ is simply the diagonal matrix containing the element-wise derivative of the virtual steering vector: $\dot{\D}_k = \mathrm{diag}(\dot{\v}_k)$. $\dot{\B}_k$ is given as
\begin{align}
[\dot{\B}_k]_{i,j} &= \dot{\v}_k^\herm\left(\G^\herm\R_\x\G \odot \Z_k^{i,j}\right)\v_k \notag\\
&+ {\v}_k^\herm\left(\G^\herm\R_\x\G \odot \Z_k^{i,j}\right)\dot{\v}_k     
\end{align}

\subsection{Derivative with Respect to Angular Spread ($\sigma_{\theta_k}$)}
Thus, $\sigma_{\theta_k}$ only influences the spreading matrix $\B_k$, leaving $\D_k$ constant. The partial derivative of the covariance matrix simplifies to:
\begin{align}
\frac{\partial \R(\tilde{\etav})}{\partial \sigma_{\theta_k}} &= q_k^2 \G \D_k \left( \frac{\partial \B_k}{\partial \sigma_{\theta_k}} \right) \D_k^\herm \G^\herm.\\
\frac{\partial[\B_k]_{i,j}}{\partial\sigma_{\sigma_{\theta_k}}}& = \v_k^\herm\left(\G^\herm\R_\x\G\odot \frac{\partial \Z_k^{i,j}}{\partial \sigma_{\theta_k}}\right)\v_k\\
\frac{\partial[\Z_k^{i,j}]_{m,n}}{\partial \sigma_{\theta_k}} &= \frac{\partial \psi_{\tilde{\theta}_{k,r}}((i-j)-(n-m))}{\partial \sigma_{\theta_k}} ,
\end{align}
where $\frac{\partial \psi_{\tilde{\theta}_{k,r}}((i-j)-(n-m))}{\partial \sigma_{\theta_k}}$ is the derivative of characterisitic function with respect to angular spread. For uniformly distributed target, this can be given as
\begin{align}
    \frac{\partial \psi_{\tilde{\theta}_{k,r}}(t)}{\partial \sigma_{\theta_k}} &= \frac{\partial \mathrm{sinc}(t\sigma_{\theta_k})}{\partial \sigma_{\theta_k}}\notag\\
    &= \frac{1}{\sigma_{\theta_k}} \left(\cos(t\sigma_{\theta_k}) - \mathrm{sinc}\left(t\sigma_{\theta_k}\right)\right)
\end{align}


	\bibliography{reference}
	\bibliographystyle{ieeetr}
    
\end{document}